\documentclass[12pt]{JHEP3}
\input epsf.tex
\epsfclipon

\usepackage{epsfig}
\usepackage{epstopdf}
\usepackage{epsf}
\input{epsf.sty}
\usepackage{graphicx,amsmath,amssymb}
\usepackage{epsfig,multicol}

\usepackage{bbm,bm,amsmath,amssymb}

\def\be{\begin{equation}}
\def\ee{\end{equation}}
\def\ba{\begin{eqnarray}}
\def\ea{\end{eqnarray}}

\newcommand{\roughly}[1]{\mathrel{\raise.3ex\hbox{$#1$\kern-0.85em
\lower1ex\hbox{$\sim$}}}}

\def\gsim{\roughly>}
\def\2pi{\left(2\pi\right)}

\def\beq{\begin{equation}}
\def\eeq{\end{equation}}
\def\beqa{\begin{eqnarray}}
\def\eeqa{\end{eqnarray}}
\def\bea{\begin{eqnarray}}
\def\eea{\end{eqnarray}}

\def\D3{\overline{\mbox{D3}}}
\newcommand{\M}{{\cal M}}

\title{On The Warped Heterotic Axion}

\author{Keshav Dasgupta, Hassan Firouzjahi, Rhiannon Gwyn\\
Ernest Rutherford Physics Building, McGill University,\\ 3600 University
Street, Montr{\'e}al QC, Canada H3A 2T8 }
\date{.. January 2006}

\abstract{We study the axion in a warped heterotic background.
It is shown that the axion decay constant, $f_a$, is sensitive to the warped mass scale of the throat.
As an explicit model, we construct a novel AdS-like geometry in heterotic string theory.
We demonstrate that in this background $f_{a}$ is given by the mass scale of the 
longest throat in the compactification. The
question of  obtaining $f_{a}$ within the required bound $10^{9}-10^{12}$ GeV is reduced to the 
construction of a throat inside the heterotic string theory compactification 
with warped mass scale in the above range. This provides a natural mechanism for realising the axion in heterotic string theory.
}

\maketitle

\begin{document}


\section{Introduction}

The axion was proposed as a plausible solution to the strong CP problem of QCD \cite{Peccei:1977ur, Peccei:1977hh, Weinberg:1977ma, Wilczek:1977pj}.  More specifically, the QCD action contains the non-perturbative term
\ba
\label{theta}
S_{\theta} = \frac{\theta}{32 \pi^{2}} \int d^{4} x \, \epsilon^{{\alpha \beta \gamma \lambda} } 
 {\bf tr} \, F_{\alpha \beta} F_{\gamma \lambda}  \, ,
\ea
where ${\bf tr} $  represents the trace in the  three-dimensional representation of $SU(3)$ and 
$\theta$ is an angular parameter. Given (\ref{theta}) one might conclude that $|\theta|$ could take any value between $0$ and 2$\pi$, but in fact there are strong observational constraints on $|\theta|$. A non-zero value of $\theta$ results in a nonvanishing dipole moment for the neutron. The recent upper bound on this dipole moment \cite{Baker:2006ts} implies that $|\theta| < 10^{-9}$.  Explaining this small value is the strong CP problem. 

The axion, $a$, is a  massless scalar field with the global shift symmetry $a \rightarrow a + \epsilon$.
This symmetry is primarily broken by QCD instanton effects with the coupling
\ba
\label{axion}
\Delta S = \frac{a}{32 \pi^{2}} \int d^{4} x \, \epsilon^{{\alpha \beta \gamma \lambda} }
{\bf tr} \,  F_{\alpha \beta} F_{\gamma \lambda}  \, .
\ea 
Combined with its kinetic energy, the action for the axion field is
\ba
\label{axion2}
S_{a}&=&  \int d^{4} x \, \left(  -\frac{f_{a}^{2}}{2} \partial_{\mu} a \partial^{\mu} a + \frac{ a}{ 32 \pi^{2}}
\epsilon^{{\alpha \beta \gamma \lambda} }  \, {\bf tr} \, F_{\alpha \beta} F_{\gamma \lambda}  \right)\nonumber\\
S_{\tilde a}&=&  \int d^{4} x \, \left( -\frac{1}{2}\partial_{\mu} \tilde a \partial^{\mu} \tilde a
+ \frac{ \tilde a}{32 \pi^{2} f_{a}}  
\epsilon^{{\alpha \beta \gamma \lambda} }  \, {\bf tr} \, F_{\alpha \beta} F_{\gamma \lambda}  \right)\, .
\ea
In the second line above the rescaled axion field $\tilde a$
is introduced such that its kinetic energy has the standard form. The parameter $f_{a}$ is called the 
axion decay constant and is inversely related to the axion mass. 

Given the similarity of (\ref{theta}) and (\ref{axion}), one can infer that the physical quantities are
independent of $\theta$. The presence of the term (\ref{theta}) can be absorbed by a constant shift in axion field via $a \rightarrow a -\theta$. Alternatively, one can think of the effect of adding the axion field into the model as
promoting $\theta$ to a dynamical field  which is naturally relaxed to zero by its vacuum expectation value.

There are strong astrophysical and cosmological bounds on the axion decay constant $f_{a}$:
\begin{equation}\label{bound}
10^{9} \, {\rm GeV} < f_{a} < 10^{12}  \, {\rm GeV} \, .
\end{equation}
If $f_a$ is less than $10^9$ GeV, the axion coupling will be too strong, leading to the production of too many axions. This would accelerate the evolution of stars such as red giants,  by transporting their energies into the outer regions more efficiently and shortening their lifetimes.
On the other hand, if $f_a$ is more than $10^{12}$ GeV, the axion coupling will be too weak, leading to the production of too much axionic dark matter in the  universe. For more details on the cosmological bounds and related  issues to do with axions see the review papers 
\cite{Khlopov, Turner:1989vc, Kim:1999ia, Sikivie:2006ni, Asztalos:2006kz} and references therein.

In this paper we would like to construct axions in models of warped heterotic string theory.
The paper is organized as follows. In Section 2 we present the construction of  the axion in warped heterotic theory, without specifying the background. In Section 3 we focus on building a new 
AdS-like warped geometry in heterotic string theory. It is shown that in this background $f_{a}$ can be lowered to values within the phenomenological window. In Section 4 we present another warped geometry in heterotic string theory where $f_{a}$ may fall within the required window. A brief conclusion and discussions are presented in Section 5.


\section{Warped Heterotic Axions }

Axions can naturally be embedded in string theory; for a review see 
\cite{Choi:1985je, Conlon:2006tq,Svrcek:2006yi, Kim:2006aq}
 and references therein.
The problem with axion construction in conventional string theory models is that it typically results in an axion decay constant higher than the range of phenomenologically allowed values. This was extensively studied in \cite{Svrcek:2006yi} with the conclusion that for string scale $m_{s}$
comparable to $M_{P}$ the axion decay constant is typically of order $10^{16}$ GeV, too big to be allowed.  

Here we consider whether it is possible to exploit the effects of warping to reduce the scale of $f_{a}$
in heterotic string theory. 
Models of warped axions in the context of a five-dimensional Randall-Sundrum scenario 
were presented in \cite{Collins:2002kp, Choi:2003wr, Flacke:2006ad, Flacke:2006re} .


\subsection{The Effect of Warping on $f_a$}
We take heterotic string theory compactified on a six-dimensional manifold ${\cal M}$, which in general will not be a Calabi-Yau (CY) manifold. Furthermore, the background spacetime is a warped geometry where the metric  is of the form
\ba
\label{metric}
ds^{2} =  h_{w}^{2} \eta_{\mu \nu} dx^{\mu} dx^{\nu} + g_{mn} (y)dy^{m} dy^{n} .
\ea
Here $h_{w}$ is the warp factor and $y^{m}, y^{n},...$ represent the coordinates along the 
manifold $\M$. The warp factor depends on the internal coordinates. Here and below the small Latin indices $m, n, ...$ represent the coordinates tangential to $\M$, the Greek indices $\mu, \nu,...$ represent the  Minkowski coordinates and the capital indices $M, N,...$ are  ten-dimensional indices.

The axion arises from the zero mode of the NS field potential $B_{MN}$ upon compactification. Depending on which component of $B_{MN}$ one considers, there are two types of axion. If all components of $B_{MN}$ are tangential to the Minkowski spacetime and are constant over $\M$, then the resulting axion is said to be ``model-independent.'' $B_{\mu \nu}$ is Hodge dual in 4 dimensions to a scalar field which is our axion. Historically, the name comes from the belief that the details of the compactification would not influence the axion construction. We shall show that this is a misnomer in the context of warped compactification since here the axion depends sensitively on the details of the compactification. On the other hand, one can consider $B_{mn}$ with components tangential to the compactified manifold ${\cal M}$. From the point of view of a four-dimensional observer this is a scalar which is then termed a ``model-dependent'' axion. 

The ten-dimensional heterotic string theory action in the Einstein frame is
\ba
\label{10D}
S=  \frac{1}{ 2 \kappa^{2}} \int d^{10} x\,  \sqrt{-g} \left( R - \frac{1}{2} \partial_{M} \phi \partial^{M} \phi-
\frac{e^{-\phi} }{2}  H^2
- \frac{\alpha'}{120} e^{-\frac{\phi}{2}}  \, {\rm Tr}~ F^{2}
\right)
\ea
where $R$ is the Ricci scalar, $\phi$ the dilaton, $F$ the $E_{8 } \times E_{8}$ or $SO(32)$ gauge field with trace in the adjoint representation of the gauge group and $H$ the NS--NS three-form constructed from the two-form potential 
$B$ as well as the curvature and the gauge field $F$.
Here $\kappa$ is the ten-dimensional gravitational coupling given by
$ 4 \pi \, \kappa^2=(2\pi \sqrt{\alpha'} )^8 \equiv m_s^{-8}$, where $m_{s}$ is defined as the string mass scale. Finally, ${\rm Tr}$ represents the trace in the adjoint representation and ${\bf tr}= {\rm Tr}/60$. 

To take into account the effect of warp factors on the graviton and NS field zero modes, we consider
dimensional reduction of each mode separately. The four-dimensional action for the zero mode of the graviton is calculated by perturbing the background in (\ref{metric}) such that  $\eta_{\mu \nu} \rightarrow \bar g_{\mu \nu}(x^{\alpha})$
where $\bar g_{\mu\nu}(x^{\alpha})$ is the metric  observed by the four-dimensional observer.
The Ricci scalar is decomposed into $R= h_{w}^{-2} \bar R + ...$ where $\bar R$ represents the Ricci
scalar constructed from $\bar g_{\mu \nu}(x^{\alpha})$.

The four-dimensional action for the zero mode of the graviton, $S_{g}^{(0)}$, is given by
\ba
S_{g}^{(0)} &=& \frac{1}{ 2 \kappa^{2}}  \int d^{4} x \, \sqrt{- \bar g} \,  \bar R  \, \int d^{6}y \,  \sqrt{g_{(6)}} \, 
  h_{w}^{2} (y) \nonumber\\
         &=& \frac{M_{P}^{2}}{2}  \int d^{4} x \, \sqrt{- \bar g} \,  \bar R, 
\ea
where $g_{(6)}$ is the determinant of the internal metric and 
\ba
\label{MP}
M_{P}^{2}= \frac{1}{ \kappa^{2}}  \int d^{6}y \,  \sqrt{ g_{(6)}} \,  h_{w}^{2} (y)
\ea
gives the Planck mass, related to the Newton constant by $8 \pi G= M_{P}^{-2}$.

To obtain the action for the zero mode of $H_{\mu \nu \lambda}$ we note that (since there is only one harmonic 
zero mode in the internal space, i.e. the constant function 1):
\ba
\label{Sg0}
 H\wedge \star H= h_{w}^{-6}   H\wedge \bar{\star} H     ,
\ea
where $\bar \star$ is constructed from $\bar g_{\mu \nu}$, independent of the warp factor.

Correspondingly, the action for the zero mode of the NS--NS field $H= H  (x^{\alpha}) $ is 
given by
\ba
\label{SH0}
S_{NS}^{(0)} =  - \frac{1}{4 \kappa^{2}} \int~ H\wedge \bar{\star} H \,
\int d^{6}y \,  \sqrt{ g_{(6)}} \,  e^{-\phi}
  h_{w}^{-2} (y)\, .
\ea

Comparing (\ref{MP}) and (\ref{SH0}), we see that in a flat background where $h_{w}= e^{-\phi}=1$, the zero modes of both the graviton and the NS--NS three-form are Planck suppressed. This is to be expected, since both of them belong to the massless sector of the closed string theory in ten dimensions.
However, as observed in \cite{Mukhopadhyaya:2002jn, Mukhopadhyaya:2007jn, Firouzjahi:2007dp},  they appear with different normalizations in a warped background.
In order to take this difference into account, we define the parameter $\beta$ such that \cite{Firouzjahi:2007dp}
\ba
\label{beta}
\beta= \frac{\int d^{6}y \,  \sqrt{ g_{(6)}} \,  e^{-\phi}
  h_{w}^{-2} (y)}{\int d^{6}y \,  \sqrt{g_{(6)}} \,  h_{w}^{2} (y)},
\ea
where the action for the zero mode of the NS--NS field is now given by
\ba
S_{NS}^{(0)}= -\frac{ \beta \, M_{P}^{2}}{4}  \int ~ H\wedge \bar{\star} H    .
\ea
The Bianchi identity for the gauge-invariant field $H$ is 
\ba \label{bianchi}
d H= \frac{\alpha'}{4} \left( {\rm tr }~ R \wedge R - \frac{1}{30}{\rm Tr}~ F \wedge F \right). 
\ea
To incorporate the axion in our construction, we dualize the $B$-field in the four-dimensional action by a scalar field $a$, via  the following Lagrange multiplier for the Bianchi identity:
\ba
\int   \, a \left[d H- \frac{\alpha'}{4} \left( {\rm tr }~ R \wedge R - {1\over 30}{\rm Tr}~ 
F \wedge F\right)\right] \, .
\ea
The action containing the $B$-field kinetic energy and the Lagrange multiplier is
\ba
\label{actionL}
S=- \frac{ \beta  M_{P}^{2}}{4} \int ~ H \wedge  \bar \star   H  +  \int ~
 a \left[d H- \frac{\alpha'}{4} \left( {\rm tr } R \wedge R 
- {1\over 30}{\rm Tr} F \wedge F \right)\right].
\ea
Integrating out $H$  in terms of the field $a$ one obtains
\ba
\bar \star H= \frac{2}{\beta M_{P}^{2}} \,  d a \,.
\ea
This is equivalent to the statement that in four dimensions the axion is Hodge dual to the anti-symmetric
$B_{\mu \nu}$ field.
Plugging this into the action (\ref{actionL}) yields
\ba
\label{axion3}
S(a)= \frac{2}{\beta M_{P}^{2}} \int d^{4} x  \, \left( -\frac{1}{2} \partial_{\mu} a  \partial^{\mu} a 
  \right) + \int a  \, \frac{\alpha'}{4} \left( {1\over 30}{\rm Tr}~ F \wedge F - {\rm tr}~ R \wedge R \right).
\ea
Upon rescaling the axion as in (\ref{axion2}) and noting that $2 \pi \sqrt {\alpha'} = m_{s}^{-1}$, we find
\ba
\label{ourf}
f_{a} = \sqrt{\frac{2}{\beta}}  \, \frac{m_{s}^{2}}{M_{P}} .
\ea
In an unwarped compactification with $\beta =1$ and taking $m_{s} /M_{P} \simeq 1/18 $ in order to get the right GUT scale from string theory, one obtains $f_{a} \simeq 10^{16}$ GeV as in \cite{Svrcek:2006yi}, too big to be acceptable. However, $\beta$ can be significantly greater than one in a warped compactification. From (\ref{ourf}) we see that this can reduce $f_a$ to the range $10^{9}-10^{12}$ GeV. In subsequent sections we will provide specific warped examples where $\beta$ is calculated to be large enough such that  
$f_{a}$ falls within the desired window.
\subsection{Example: Heterotic Compactification on a non-K\"ahler Manifold}
There have been many attempts to embed an axion construction in string theory (for an extensive review see
\cite{Svrcek:2006yi} and references therein). Starting with  a string scale
comparable to $M_{P}$, it can be shown \cite{Svrcek:2006yi} that in most 
cases $f_{a}$ is close to the GUT scale.
The exceptions are when the Standard Model gauge fields are supported on vanishing cycles, where
it is possible to lower $f_{a}$ to the desired phenomenological bound \cite{Svrcek:2006yi}.

Axion construction in very large compactification volumes was studied in \cite{Conlon:2006tq}. 
A very large compactification corresponds to a low-scale string theory. It is argued that up to
numerical factors of order unity, $f_{a} \sim m_{s} \sim 10^{11}$ GeV. 

The axion in a warped heterotic background was considered in \cite{Kim:2006aq}. The model considered there is a heterotic
compactification on a non-K\"ahler manifold \cite{sav, beckerD, bbdg, GP, bbdgs, bbdp, gttwo}. The non-K\"ahler background
is a non-trivial $T^2$ fibration over a K3 base. In the Einstein frame the full ten-dimensional metric can be written in the
following way:
\begin{equation}\label{nkmanifold}
ds^2 = e^{-{\phi\over 2}} \eta_{\mu\nu} dx^\mu dx^\nu + e^{-{\phi\over 2}}\left[(dx+\alpha_{1})^2 + (dy+\alpha_{2})^2\right]
+ e^{{3\phi\over 2}} ds^2_{\rm K3},
\end{equation}
where $x$ and $y$ are local coordinates such that $dx\,+\,idy$ is a holomorphic form on the $T^2$ fibers, and the
$\alpha_{i}$ are local one-forms on the K3 base.  
For this particular compactification we see that the dilaton is related to the warp factor
via $e^{-\phi} = h_{w}^{4}$. Plugging this into our expression for $\beta$ in (\ref{beta}), one finds
that $\beta=1$. As mentioned in \cite{Kim:2006aq}, the warping does not help to reduce $f_{a}$
for the model-independent axion in the above background. 
This indicates that the axion decay constant is large
if one starts with a large string scale.
One way out of this conclusion is to construct a background where the dilaton is independent
of the warp factor, so $\beta$ can be made sufficiently large. The backgrounds studied in
Sections 3 and 4 of the present work both satisfy this condition and we bypass the difficulty of cancellation of
the warp factor in (\ref{beta}) mentioned above.

\subsection{Example: ``Model-Dependent" Axions in IIB}
In this work we focus on the construction of  a ``model-independent'' axion. However, one may naturally
ask whether ``model-dependent'' axions with the allowed decay constants  can be constructed in our formalism. 
The answer seems to be affirmative\footnote{We thank Joe Conlon for discussion on this issue.}, although
in this paper we will only address this issue briefly 
because it requires us to know the details of the cohomological and 
homological properties of the internal space. 

To give one example, 
consider a model where the axion arises from the zero mode of the RR four-form potential $C_{(4)}$
in type IIB string theory, say for example the background \cite{ouyang}. 
To support the axion we need a D7-brane such that
\ba
\int C_{(4)} \wedge F\wedge F
\ea
is non-zero, where the integration is over the D7 worldvolume. It is assumed that  $F$ has legs along
the Minkowski coordinates, while $C_{(4)}(x^{\alpha}, y^m)$ 
has legs entirely along the compactified directions with $x^\alpha$ and $y^m$ denote the coordinates on the 
Minkowski and the internal spaces respectively.  
This means that we are decomposing $C_4$ as 
\begin{equation}\label{cfour}
C_4 (x^\alpha, y^m) ~ = ~ \varphi(x^\alpha) \otimes h_4(y^m)
\end{equation}
where $h_4(y^m)$ is a harmonic four-form in the internal space and $\varphi(x^\alpha)$ is a scalar which will 
have axion-like couplings. Clearly, since the harmonic four-forms in the internal space are classified by the 
second Betti numbers $b_2$, there are $b_2$ axions from this decomposition. 
In the following we will choose $b_2 = 1$ to get a 
single axion for our case, but this is of course a model-dependent statement. 
Furthermore, it is also easy to see that in terms of powers of the warp factor $h_w$, the kinetic energy
of the axion scales like
\ba
\int \sqrt{g_{(6)}} \,  h_{w}^{2} \vert h_{4}\vert^{2} \, .
\ea
where $\vert h_{4}\vert$ is the magnitude of the harmonic four-form in the internal space. Using Hodge duality, this
could be mapped to the harmonic (1, 1) form, $h_2$, in the internal space. 
Now, constructing $\beta$ using (\ref{beta}) we see that:
\begin{equation}\label{betnow}
\beta= \frac{\int d^{6}y \,  \sqrt{ g_{(6)}} \,
  h_{w}^{2} (y) \vert h_2\vert^2}{\int d^{6}y \,  \sqrt{g_{(6)}} \,  h_{w}^{2} (y)},
\end{equation}
which may be significantly bigger than one depending on the behavior of the harmonic two-form in the internal space. 
Unfortunately, the exact form of $h_2$ for a CY (or non-CY) manifold is not known so we cannot make 
any concrete statement here.\footnote{To determine the harmonic form we need the metric of the internal space (say 
CY). So far there is no known solution for the metrics of compact CY spaces.}
Secondly, we require the harmonic form to be peaked near the throat of our internal
space, which again requires us to know the precise form for the $h_2$. 
We do believe however that there is no conceptual problem in finding some internal space that can give rise to the 
required $h_2$ forms.
  
This conclusion can also be generalized to the zero modes of other ``model-dependent'' axions. The upshot is
that model-dependent axions look like scalars from a four-dimensonal point of view and, as one can easily check,  
$\beta$ for a scalar is dependent on the warp factors as well as
the magnitude of the harmonic forms in the internal space. Tuning these correctly we can possibly have a large 
$\beta$ from such compactifications although the results depends crucially on our knowledge of these harmonic forms (which 
is lacking at this stage). 

The model-independent axion on the other hand 
is an exception: we are not required to know the detailed topological properties of the internal space as 
there is one and only one harmonic zero form in the internal space (which can be set to 1). Additionally, 
as mentioned in  \cite{Mukhopadhyaya:2002jn, Mukhopadhyaya:2007jn},  tensors of rank three and higher 
have the interesting property that their zero
modes are strongly suppressed compared to the graviton zero mode.

Our goal in the rest of this paper is to construct warped geometries in heterotic string theory to
see whether $\beta$ for the model-independent axion $B_{\mu \nu}$
can be made sufficiently large to lower $f_{a}$ into the narrow phenomenological window.


\section{An AdS-type Background in Heterotic Theory}
\label{AdS}
As explained above, in order to get large enough $\beta$ we need to construct warped geometries where the dilaton is independent of the warp factor. We construct backgrounds with precise warp factors using certain identifications between 
($p, q$) and ($0, q$) sigma models where $p = q = 1, 2, 4$ for bulk ${\cal N} = 1, 2, 4$ 
supersymmetries respectively. 
A class of the resulting heterotic backgrounds will resemble warped AdS$_5$ backgrounds.
In the following we will discuss the sigma model identifications that we shall use 
to construct our backgrounds. Readers interested only in the final result should skip Section \ref{sigmamodel} and go directly to Section \ref{construction} where our backgrounds are presented. In Section 4 we present another new heterotic background which has a warp-independent dilaton but non-trivial torsion. 

\subsection{Sigma-Model Constructions}
\label{sigmamodel}
Our present analysis will require us to study non-linear sigma models in both type II and heterotic theories,
so we begin with a brief review of the method. First note that there are two ways to drag a type IIB background to the heterotic side:

\vskip.1in

\noindent $\bullet$ $U$-dualise a type IIB toroidal orientifold background to heterotic, or

\noindent $\bullet$ use sigma model identification to bring a torsional type IIB background to heterotic.

\vskip.1in

\noindent The two techniques achieve similar goals, but each comes with distinct advantages and disadvantages. To begin with, the first technique demands a ten-dimensional framework while the second works only from the two-dimensional point of view. Secondly, both techniques will in general involve some kind of U-duality needed to bring the type IIB theory into the required form. For the first, one must find an orientifold limit of a given type IIB background. This is not always easy. Being able to lift a given type IIB background to F-theory in principle guarantees the existence of an orientifold point but the lifting is not always easy to realise in practice. Orientifolding in the presence of background
fluxes implies projecting out some components of the fluxes, and therefore finding a consistent orientifold corner 
of a given F-theory compactification can turn out to be subtle.  

The second technique might at first appear easier than the first, but this is not always the case since not every type IIB background can be pulled to the heterotic side. In particular, only specific choices of IIB fluxes are allowed by such an operation, so the presence of background fluxes can prove an obstacle. Since a generic warped type IIB compactification may therefore not have a heterotic dual in the latter, i.e. sigma model sense, it is sometimes necessary to modify the type IIB background in order to be able to find a heterotic dual. In general, a judicious choice of the appropriate technique should allow us to find the required heterotic dual from type IIB theory. 

Our aim in the following analysis is to develop 
a heterotic background with say ($0, p$) worldsheet
supersymmetry from a given type IIB background with ($p, p$) worldsheet supersymmetry. 
The simplest way to do this is to add 
non--interacting fields to the sigma-model action with ($p, p$) supersymmetry. In general this will ruin the
carefully balanced ($p, p$) supersymmetry of this model. We can
use this to our advantage by adding non--interacting fields {\it
only} in the left--moving sector. This breaks the
left--moving supersymmetry, and one can therefore hope
to obtain an action for a ($0, p$) model from the ($p, p$) model. 

To make this precise, let us consider the simplest case with $p = 2$ 
and define the corresponding sigma--model action in the following way:\footnote{For $p =1$ the situation is a little
subtle as we will discuss later.}
\begin{equation} \label{supn}
S = \frac{1}{8\pi \alpha'} \int d^2\sigma \Big[(g_{ij} +
B_{ij}) \partial_+ X^i \partial_- X^j + {1\over 4}S^g_{\rm fermionic}\Big],
\end{equation}
where the $X^i$ are the bosons and $S^g_{\rm fermionic}$ denotes the fermionic part of the action.
Written this way, the action requires no other corrections\footnote{For example Chern-Simons corrections.} and will consequently be anomaly free. The fermionic part is made of a right--moving sector containing eight fermions (which we denote as $\psi^p$) and a left--moving sector also containing eight fermions (which we denote as $\psi^{\dot q}$). Together they give rise to the type II worldsheet action.

Taking into account the NS three-form fields $H_{NS}$, the fermionic part of a Green-Schwarz 
superstring can be written as 
\begin{equation}\label{gsstring} 
S^g_{\rm fermionic} = 4i \psi^p \Delta_+ \psi^p +
4i \psi^{\dot q} \Delta_- \psi^{\dot q} + R_{(+)ijkl}
\sigma^{ij}_{\dot p \dot q} \sigma^{kl}_{rs}\psi^{\dot p}
\psi^{\dot q} \psi^{r}\psi^{s},
\end{equation}
where $\psi^{p}$ and $\psi^{\dot q}$
are the two inequivalent spinor representations of the transverse
$D_4$ and the sigma matrices are defined as $\sigma^{ij}_{\dot p
\dot q} \equiv \Gamma^{[i}_{r[\dot p}\Gamma^{j]}_{\dot q]r}$ with
a similar definition for the other components. The Gamma matrix
has $8 \times  8$ blocks given as 
\begin{equation}\label{gamaram}
\Gamma^i \equiv  \begin{pmatrix} 0&
\Gamma^i_{p\dot q}\\ \Gamma^i_{\dot r s}& 0 \end{pmatrix},
\end{equation}
which are used to
define the $\sigma$s above. In this notation we have to specify
what we mean by the covariant derivative $\Delta_\pm$ and the curvature
$R_{(+)ijkl}$. The most generic definition of the covariant
derivative is given by \cite{hull,bbdg}
\begin{equation}\label{lapid}
\Delta_\pm
\psi^{q(\dot q)} = \partial_\pm \psi^{q(\dot q)} + \frac{1}{2}
\left(\omega - {1\over 2} H \right)^{ab} \sigma^{pq(\dot p \dot
q)}_{ab} \psi^{p(\dot p)},
\end{equation}
where ${1\over 2} H$ is the torsion and
we have chosen the torsional connection $\omega_+ \equiv \omega -
{1\over 2} H$, and not $\omega_- \equiv \omega - {1 \over 2} H$ (see \cite{bbdg} for
details on this). Observe that in the absence of $H_{NS}$ there is
no such ambiguity in the definition of the GS superstring. Finally, the curvature $R_{(+)ijkl}$ that we defined above is 
measured w.r.t. the connection $\omega_+$, where
\begin{equation}
R^{i}_{(+)jkl} = \partial_k \Gamma^{(+)i}_{lj} +
\Gamma^{(+)i}_{km}\Gamma^{(+)m}_{lj} - (k \leftrightarrow l),
\end{equation}
and $\Gamma^{(+)i}_{jk}, H^i_{jk}$ are defined as
\begin{equation}
\Gamma^{(+)i}_{jk} \equiv \Big\{{}^i_{jk}\Big\} + {1\over 2} H^i_{jk}, ~~~~H^i_{jk} = g^{il} H_{jkl}.
\end{equation}

The above therefore gives the classical action for the type II string with background NS three-form fluxes. 
To get the (0, 2) heterotic action from the type II action described above, we take the following steps:

\vskip.1in

\noindent $\bullet$ Keep the right--moving sector unchanged, i.e. the $\psi^p$ remain as before.

\noindent $\bullet$ In the left--moving sector, replace the $\psi^{\dot q}$ by eight
fermions $\Psi^A$, $A = 1, ... 8$. Also add 24 additional non--interacting
fermions $\Psi^B$, $B = 9, ... 32$.

\noindent $\bullet$ Replace $\omega_+$ by the gauge field $A$, i.e. embed the torsional
spin connection into the gauge connection.

\vskip.1in

The above set of transformations will convert the classical type II
action given in \eqref{supn} to a classical heterotic one i.e. (0, 2) one. One might,
however, wonder about the Bianchi identity in the heterotic theory.
The type IIB three--form fields are closed, whereas heterotic
three--form fields satisfy the Bianchi identity. These statements seem to be reconciled by the embedding 
$\omega_+ = A$, which should result in a closed heterotic three-form. However, because
of subtleties discussed in \cite{bbdg, smit, papad},  the above
embedding will not allow any compact non--K\"ahler manifolds in
the heterotic theory, so this embedding is not an admissible solution to the problem. Therefore, as a first approximation,
we assume an embedding of the
form 
\begin{equation}\label{bedd}
\omega_+ ~ = ~ A ~ + ~{\cal O}(\alpha').
\end{equation}
Using this, the new action with (0, 2) supersymmetry becomes
\begin{eqnarray}\label{snow} \nonumber
S &=& {1\over 8\pi \alpha'} \int d^2\sigma
\Big[(g_{ij} + B_{ij}) \partial_+ X^i \partial_- X^j + i \psi^p (\Delta_+
\psi)^p + i \Psi^A (\Delta_-\Psi)^A + \\ &&~~~~~~~~~~~~~~~~~ + {1\over
2}F_{ij(AB)} \sigma^{ij}_{pq} ~\psi^p \psi^q \Psi^A \Psi^B + {\cal
O}(\alpha')\Big],
\end{eqnarray}
 where due to the embedding \eqref{bedd}, $F^a_{ij}$ is the Yang--Mills field strength measured w.r.t. Lie algebra
matrices $T^a_{AB}$. The fermion indices are $A = 1,..., 32$, which means
there are 32 fermions, and hence the $T^a$ form tensors of
rank 16. The reader can easily identify the above action as an
action for a heterotic sigma model with torsion \cite{hull, bbdg, bbdgs,
rstrom, gross, hulltown}. The action of the
Laplacian \eqref{lapid} changes accordingly to
\begin{eqnarray}\label{lapidnow} \nonumber
 &&\Delta_-\Psi^A  ~ =~  \partial_-\Psi^A + A^{AB}_i
(\partial_-X^i)~\Psi^B;\\  && \Delta_+\psi^p  ~ = ~ \partial_+\psi^p + {1\over
2}(\omega_+)^{ab} \sigma^{pq}_{ab} \psi^q; \\ \nonumber  && H_{ijk} ~ = ~  {1\over
2}\left(B_{ij,k} + B_{jk,i} + B_{kj,i}\right).
\end{eqnarray}
At this stage the replacements \eqref{lapidnow} may lead us to think that the sigma model has only 
(0, 1) supersymmetry. This is only superficial, and the situation is  
similar to the (1, 1) action for the type
II case \cite{hull}. The full (0, 2) susy will be determined by additional
actions on the fields (exactly as for the (1, 1) case). In the absence of torsion the above action \eqref{snow} 
takes the following familar form:
\begin{equation}\label{knosnow}
S = {1\over 8\pi \alpha'} \int d^2\sigma \left[\delta_{ij} \partial_+ X^i \partial_- X^j + i \psi^p \partial_+ \psi^p
+ i \Psi^A \partial_- \Psi^A\right],
\end{equation}
with the equations of motion
\begin{equation}
\partial_+ \psi^p = 0; ~~~~~~~~ \partial_- \Psi^A = 0,
\end{equation}
that give us the orientations of the two worldsheet fermions. 

Classically the action \eqref{snow} is invariant under the following transformations:
\begin{eqnarray}\label{transfo} \nonumber 
&& \delta \Psi^A = \Lambda^A_B(X) \Psi^B, ~~~~~~~~~~~~~ \delta A^{AB}_i = \partial_i \Lambda^{AB} + 2 A_i^{C[A}\Lambda^{B]C},\\
&& \delta \Psi^P = {1\over 2} \Theta^{ab}(X) \sigma^{pq}_{ab} \Psi^q, ~~~~~~~ \delta\omega_i^{ab} = \partial_i \Theta^{ab} 
+ 2 \omega_i^{c[a}\Theta^{b]c},
\end{eqnarray}
where $\Lambda$ denotes the local group rotations and $\Theta$ denotes local SO(8) Lorentz rotations. However, as
is well known, these transformations are anomalous. The anomaly can be cancelled by the following choice of our
three-form flux:
\begin{equation}\label{torfor}
H = dB - \alpha'\Big[{1\over 30} {\rm Tr}\Big(A \wedge F - {1\over 3} A \wedge A \wedge A\Big) - {\rm tr} 
\Big(\omega_+ \wedge R_{(+)} - {1\over 3}\omega_+ \wedge \omega_+ \wedge \omega_+\Big)\Big],
\end{equation}
which is of course the correct description for the torsional three-form $H$ in the heterotic theory. Observe that 
the torsion $H$ appears on {\it both} sides of the equation \eqref{torfor}.  
 
The above set of manipulations
that convert a classical type II action to a classical heterotic
one will help us to understand various things about the heterotic theory using data of type
II theories. One important thing to notice is that 
both the metric and the $B_{ij}$ fields can be taken to the
heterotic side provided the original
type II background does not have any $H_{RR}$ fields. In the presence
of $H_{RR}$ the simple manipulations that we performed cannot give
a (0, 2) or a (0, 1) model. We are therefore particularly interested in
type II models that allow only for an NS three-form.
When both NS and RR backgrounds are present, it might still be possible to perform the above 
manipulations if one can find an equivalent U--dual background. This U--dual background will in general not be K\"ahler
(not even conformally K\"ahler). Observe that these U-dualities do not
require the original background to be at the orientifold point.
This is therefore different from the analysis performed in \cite{sav,
beckerD}.\footnote{Another interesting question would be to allow for both
NS and RR backgrounds in the S--dual type I picture. This is a
highly restrictive scenario as the allowed values of NS fluxes in
the S--dual type I picture leave us with only two discrete choices
\cite{sensethi}.}

Yet another issue is the existence of the bundle. This can be inferred from the
detailed analysis given (at least for the $U(1)$ case) in \cite{bbdgs, lust}. However, the
situation here may become a little simpler than the one in \cite{bbdgs}, because
of the embedding \eqref{bedd}. Recall that for the manifold studied in \cite{bbdgs}, a given complex structure $J$ must satisfy the following constraint
\begin{equation}\label{comp}
i \partial\bar\partial J ~ = ~ 0 + {\cal O}(\alpha'),
\end{equation}
which means that we can study the stability of bundles using the
recent analysis of Li and Yau \cite{yauli} and Fu and Yau \cite{yaufu}. For our case we will analyse the vector 
bundle directly from the sigma model by coupling the closed string worldsheets to open string worldsheets. 
We will touch on this issue later in the
paper. 

To construct an explicit heterotic background we have to make sure that the susy variations of the gravitino 
field $\chi_i$ and the gaugino field $\lambda^A$ vanish in this background under an arbitrary susy parameter 
$\eta(X)$. This translates to the following set of well-known conditions:
\begin{eqnarray}\label{lbaaz} \nonumber
 \delta \chi_i & \equiv & \partial_i\eta + {1\over 2} \Big(\omega_i^{ab} + {1\over 2} H_i^{ab}\Big)\Gamma_{[a}\Gamma_{b]}
\eta ~ = 0, \\
 \delta \lambda^A & \equiv & F^A_{ij} \Gamma^{[i}\Gamma^{j]} ~ = ~ 0.
\end{eqnarray}
Therefore to conclude this section we see that we can drag a given type IIB background to the heterotic side 
using our sigma-model identifications provided under U-dualities the IIB background has a non-trivial NS three-form 
and a metric with no RR forms. In the following section we will provide concrete constructions of a class of 
such backgrounds. 


\subsection{Construction of New Heterotic Backgrounds}
\label{construction}
From Section \ref{sigmamodel} we know that we require a type IIB background with non-trivial metric and NS three-form. We also 
require a dilaton independent of the radial coordinate\footnote{The dilaton will in general not be a constant in the heterotic theory because of the presence of torsion. The only exception is when the torsion is generated by 
Chern classes as we shall see in Section 4.}
so that $\beta$ can be made reasonably large. One background
that immediately comes to mind is type IIB theory on $AdS_5$ space. However, the minimally supersymmetric $AdS_5$ background, 
i.e. $AdS_5 \times T^{1,1}$,  given by Klebanov-Witten \cite{klebwit} cannot be easily pulled to the heterotic side using our sigma--model identification. The non-trivial fibrations of the internal space $T^{1,1}$ create extra fluxes under U-duality which prohibit a heterotic dual for this background.  We are therefore
left with the other choices: $AdS_5 \times S^5$ and 
$AdS_5 \times {S^5\over {\bf Z}_n}$ with\footnote{As is well known, the $n = 2$ case is the simplest, with the ${\bf Z}_2$ 
action considered in a specific way \cite{klebwit}. In general the typical ${\bf Z}_n$ action involves extra 
seven branes in the picture \cite{fayyaz1, fayyaz2}. We will discuss this later.} $n = 2, 3, 4, 6$. 

Using our analysis in the previous section we now claim that in the heterotic theory we will have a background of the 
form
\begin{equation}\label{nhebag}
ds^2 = e^\phi ds^2_{AdS_5} ~ + ~ ds^2_{X^5},
\end{equation}
that satisfies all the requirements laid down in the previous section. Here 
$\phi$ is the dilaton that depends only on the coordinates of the internal space $X^5$ and not on the radial
coordinate $r$. This non-trivial dilaton will be supported by a background torsion $H$. 


To be specific, our earlier arguments then require us to 
start with an $AdS_5 \times S^5$ background in type IIB string theory given (in units of $\alpha'$) by
\ba
\label{Ads5}
ds^{2}= \frac{r^{2}}{R^{2}} \eta_{\mu \nu} dx^{\mu} dx^{\nu} + \frac{R^{2}}{r^{2}}~dr^2
+ R^{2} d\Omega_{5}^{2},
\ea
where $\mu, \nu = 0, 1, 2, 3$ are the spacetime directions and $R$ is the curvature radius of the AdS space given by
\ba
\label{R}
R^{4} = 4 \pi g_{s} N \, .
\ea
Here $N$ is the quantised charge of the five-form $F_{5}$:
\ba
\int_{S^{5}} F_{5} = (4 \pi^{2} \alpha')^{2} \, N \, .
\ea
Furthermore, in the absence of NS and RR three-forms the five-form $F_5$ can be written as
$F_{5}= d C_{(4)} + * d C_{(4)}$ with the RR four-form, $C_{(4)}$, given by
\ba
\label{C4}
C_{(4)} = \frac{r^{4}}{   g_{s} R^{4}} \,  dx^{0} \wedge dx^{1} \wedge dx^{2} \wedge dx^{3} \, .
\ea
Finally the metric of the five-sphere,  $d\Omega_{5}^{2}$, in (\ref{Ads5}) is 
\begin{equation}
\label{sfive}
d\Omega_5^2 = d\gamma^2 + {\rm cos}^2 \gamma ~d\varphi_3^2 + {\rm sin}^2\gamma 
\left(d\psi^2 + {\rm cos}^2 \psi ~d\varphi_1^2 + {\rm sin}^2\psi ~d\varphi_2^2\right) \, ,
\end{equation}
where $0 \leq \gamma, \psi \leq \frac{\pi}{2}$ and $ 0 \leq \phi_i \leq 2 \pi$.
We see that there are three local isometries along directions $\varphi_1$, $\varphi_2$ and 
$\varphi_3$. We can choose $\varphi_1$ and $\varphi_2$ as the directions along which to perform our T-dualities, but we have to take care because there are no global one-cycles in the manifold. In fact, at the points
\begin{equation}\label{ncpoint}
\gamma ~ = ~ 0; ~~~~ \psi ~ = ~ 0; ~~~~ \psi ~ = ~ \frac{\pi}{2},
\end{equation}
the cycles all shrink to zero size and the U-dual manifold will be non-compact. To avoid these issues, we
will make our U-dualities away from the points \eqref{ncpoint}. Now using the sigma model identifications, we find the  
following background in heterotic theory:
\begin{eqnarray}\label{hetbag}
\nonumber ds^2 &=& {1\over 2} R^2  {\rm sin}^2 \gamma ~{\rm sin}2\psi \Big[{r^2 \over R^2} 
dx_\mu dx^{\mu} + {R^2 \over r^2} ~dr^2\Big]  \\
 \nonumber &+& {1\over 2} R^4 {\rm sin}^2 \gamma ~{\rm sin}2\psi \left[d\gamma^2 + {\rm cos}^2 
 \gamma d\varphi_3^2 
+ {\rm sin}^2 \gamma~d\psi^2\right] + {\rm tan}\psi \, d\varphi_1^2 + {\rm cot}\psi \, d\varphi_2^2 ; \\
 e^\phi~ &=& {1\over 2g_s}\left(R^2 ~{\rm sin}^2 \gamma ~{\rm sin}~2\psi\right)  \, ;  \\
 \nonumber H &=& {4r^3 \over R^4}  e^{2 \phi}~\ast \left(dx^0 \wedge dx^1 \wedge dx^2 \wedge dx^3 \wedge dr \wedge d\varphi_1  
\wedge d\varphi_2\right) + {\cal O}(\alpha')  \\
\nonumber &=& -\frac{4 R^{4}}{g_s^2} \sin^{3} \gamma \cos \gamma \sin \psi \cos \psi \,    d \gamma \wedge  d\psi 
\wedge  d \varphi_{3}      + {\cal O}(\alpha')        \,  ,
\end{eqnarray}
with an additional vector bundle that we will describe later. This vector bundle has to satisfy the modified 
DUY equations which appear because of the background torsion \cite{bbdg, bbdgs}. Furthermore our Hodge star operation
is defined for a generic $p$-form in the following way:
\begin{equation}\label{hodge}
\left(\ast \omega\right)_{\mu_1 \mu_2 ..... \mu_{10-p}} = \frac{\sqrt{-g}}{p!} ~
\epsilon_{\mu_1 \mu_2 ..... \mu_{10-p}}^{~~~~~~~~~~~~~~\nu_1 \nu_2 .....\nu_p} 
\omega_{\nu_1 \nu_2 .....\nu_p}.
\end{equation} 
Observe that the new background on the heterotic side is not quite an $AdS_5$ background because of the unusual 
warp factors, although the radial dependence remains like that of the standard type IIB $AdS_5$ background. The internal 
space is also not an $S^5$ anymore. The metric has non-trivial warp factors that make the background non-K\"ahler. 
Furthermore, the dilaton is not a constant, and $H$ is in general more complicated than the 
standard form\footnote{Recall that in the derivation of the dimensional reduction of a ten-dimensional
heterotic action on a manifold with torsion \cite{lustu}
it is assumed that the warp factor is exactly equal to the dilaton (as in \cite{rstrom} for example). 
For us this is not the case, so there will 
be corrections. 
Additionally 
our background \eqref{hetbag} does not have the standard form of G-structures discussed in \cite{gauntlett}. 
We will address these issues elsewhere.}
\begin{equation}\label{hthree}
H = e^{2\phi} \ast\left[d(e^{-2\phi} J)\right].
\end{equation}
However we do expect the anomaly-cancelling Bianchi identity with $R_{(+)}$ and a vector bundle with curvature $F$
defined on the internal six-dimensional space, to hold as: 
\ba
dH ~ = ~ \frac{\alpha'}{4} \left( {\rm tr }~ R_{(+)} \wedge R_{(+)} - {1\over 30}{\rm Tr}~ 
F \wedge F\right).
\ea
To read off physical quantities, we transform the metric into the Einstein frame via
$g_{MN}^{(E)}= e^{-\phi/2} g_{MN}^{(S)}$. After restoring the necessary factors of $\alpha'$ and rescaling $x^\mu$ 
($ g_{s}^{1/4} \alpha'^{1/2}\, x^{\mu} \rightarrow x^{\mu}$), \eqref{hetbag} 
in the Einstein frame is given by
\ba 
\label{Ein1}
ds^{2}&=& \sin \gamma \, \sqrt{\sin \psi \cos \psi} 
\Bigg[  \frac{r^2}{R } dx^{\mu} dx_{\mu}   + \alpha' \sqrt{g_s} R^{3} \left( 
 \frac{dr^{2}}{r^{2}}  + d \gamma^{2} + \cos^{2} \gamma d \phi_{3}^{2} + 
\sin^{2} \gamma d \psi^{2} \right)\Bigg]  \nonumber\\
&&+ \frac{\alpha' \sqrt{g_s}  \sin^{\frac{1}{2}} \psi  }{R \cos^{\frac{3}{2}} \psi   \sin \gamma  } d\phi_{1}^{2}
+\frac{  \alpha'\sqrt{g_s} \cos^{\frac{1}{2}} \psi  }{R \sin^{\frac{3}{2}} \psi   \sin \gamma  } d\phi_{2}^{2},
\ea 
with $H$ given as in \eqref{hetbag}. Note that this geometry has the form of a warped metric (\ref{metric})
with warp factor 
\ba\label{wfac}
h_{w}^{2} = \sin \gamma \sqrt{\sin \psi \cos \psi}  \,  \frac{r^2}{R }.
\ea
One can check that the background given by (\ref{Ein1}) or equivalently \eqref{hetbag}
is a consistent solution.  With $H$ given as in \eqref{hetbag}, the equation of motion 
$d \star H = 0$  is trivially satisfied. The dilaton equation,
\ba
\label{dileom}
\frac{1}{ \sqrt{-g} } \partial_{M} \left(   \sqrt{-g}     \partial^{M} \phi  \right) + \frac{e^{-\phi} }{12} H^{2}=0,
\ea
is also satisfied. Finally the Einstein equation, $G_{MN}= \frac{1}{2} T_{MN}$, where $G_{MN}$
is the Einstein tensor and $T_{MN}$ is the stress-energy tensor given by
\ba
\label{Tmn}
T_{MN}=  \partial_{M} \phi \,   \partial_{N} \phi  
- \frac{1}{2} g_{MN} \partial_{P} \phi  \,  \partial^{P} \phi 
+ \frac{e^{-\phi} }{2 }  H_{MPQ} H_{N}^{\, \, PQ }
-\frac{e^{-\phi}}{ 12}   g_{MN}   H^2,
\ea
is also satisfied.\footnote{More details on how this solution solves the Einstein equations are given in the Appendix.} This gives us a powerful test of the consistency of our background. As mentioned above, the background 
\eqref{hetbag} is well defined away from the points \eqref{ncpoint}. Our sigma-model identification clearly fails 
at these points. It is then no surprise that the Ricci scalar for the metric (\ref{Ein1}), given by 
\ba
{\cal R }= \frac{  1+ 4 \sin^{2} \psi  \cos^{2} \psi \sin^{2} \gamma}{2 \alpha' \sqrt {g_s}    R^{3} \sin^{3} \gamma
 \sin^{5/2} \psi   \cos^{5/2} \psi       } \, ,
\ea
diverges at the points \eqref{ncpoint} mentioned above. 

Such divergences can easily be cured by removing these points from the original $S^5$ \eqref{sfive}. Then the metric \eqref{hetbag} is a good description of the geometry away from these points, and the global six-dimensional manifold will be a compact non-K\"ahler manifold when we cut off the radial direction and replace it with a smooth cap. Physically, the quantity $g_sN$ corresponds to the integral of the three-form over a three-cycle of our non-K\"ahler manifold. As mentioned above, the four-dimensional spacetime is then a warped Minkowski spacetime with warp factor given by \eqref{wfac}. 

Once the singular points are smoothed out, the manifold will have a well-defined Riemann tensor globally. In fact there will now be an ${\cal O}(\alpha')$ correction to the torsion $H$ \eqref{hetbag} from the Riemann tensor two-form 
${\cal R}_\omega$ that is linear in the torsion one-form. The total torsion ${\cal H}$ 
for the smooth manifold will then be a 
combination of the result \eqref{hetbag} and the new contribution from the two-form ${\cal R}_\omega$. This is  
given by:
\begin{equation}\label{tors}
{\cal H} ~ \equiv ~ H  +  \frac{\alpha'}{2}~{\rm Tr}\left({\tilde H}_{(3)} \wedge {\cal R}_\omega  +  ....\right)
+ {\cal O}(\alpha'^2),
\end{equation}
where $H$ is defined in \eqref{hetbag}, ${\tilde H}_{(3)}$ is the one-form created from $H$ using the vielbeins and ${\cal R}_\omega$ is defined, using the gravitational spin-connection at zeroth order in $\alpha'$, 
in the following way:
\begin{equation}\label{romega}
{\cal R}_\omega ~ = ~ d\omega + \frac{2}{3} \omega \wedge \omega.
\end{equation}
Thus the trace in \eqref{tors} is the natural trace over the holonomy group of the internal manifold. The dotted 
terms involve higher orders in ${\tilde H}_{(3)}$ whose discussion we postpone for the next section. Here we want to point
out that the total torsion \eqref{tors} is well defined for our non-K\"ahler manifold as long as ${\cal R}_\omega$ is 
well defined. It is also not closed, and satisfies
\begin{equation}\label{binc}
d{\cal H} ~ = ~ {\cal O}(\alpha').
\end{equation}
This ${\cal O}(\alpha')$ can come from two potential sources: from the curvature correction discussed above, and from the embedding \eqref{bedd}. We expect these contributions to combine nicely to give us the expected Bianchi 
identity.
\subsubsection{The Vector Bundle}
As emphasised before, the embedding \eqref{bedd} implies that the spin connection cannot be embedded 
in the gauge connection
to get a full non-K\"ahler geometry. So the question is to determine the precise vector bundle on the heterotic
side. Using our sigma-model identification, all we now need is to add open string worldsheets. 
This is subtle because generic open string worldsheets would ruin the conformal invariance in the original
type IIB picture. 
Now, parametrising a unit radius $S^5$ in type IIB by
\begin{equation}\label{para}
z ~=~ {\rm cos}~\gamma ~e^{i\varphi_3}, ~~~~~ v ~=~ {\rm sin}~\gamma ~{\rm cos}~\psi~e^{i\varphi_1}, ~~~~~
{\tilde v} ~=~ {\rm sin}~\gamma ~{\rm sin}~\psi~e^{i\varphi_2},
\end{equation}
we see that the $S^5$ can be written as
\begin{equation}\label{sfn}
 d\Omega_5^2 = \vert dz\vert^2 + \vert dv\vert^2 + \vert d{\tilde v}\vert^2,
\end{equation}
{}from which it is clear that vector bundles in the heterotic side should correspond to seven-brane configurations 
that are stretched along $z$ and wrap a non-trivial two-cycle in ($v, {\tilde v}$) space. This configuration
should have no global charge, and the distances between the seven branes should be parametrised by the ($\varphi_1, \varphi_2$) coordinates. It is possible to preserve conformal invariance in this set-up, so that the IIB 
metric remains $AdS_5$ and the global symmetries are $G_i$. 
Under U-duality this will then give us the vector bundles in the heterotic side. 
   
With these constraints the background \eqref{hetbag} will allow the corresponding $G_i$ vector bundles
respectively. The original $SO(32)$ or $E_8 \times E_8$ gauge symmetry of the heterotic string
would then be broken to any of these  
gauge symmetries by the corresponding Wilson lines.

On the other hand, if the seven branes wrap a four-cycle in ($v, \tilde v$) space instead of a two-cycle
there is a simple way to get conformally invariant backgrounds with open strings: these are the 
constant coupling backgrounds of \cite{senF, DMF}, whose AdS limits were worked out in \cite{fayyaz1, fayyaz2}. 
In terms of the bulk picture this is an F-theory compactification with 
appropriate seven branes inserted \cite{vafaF}. Arranging these seven branes in certain special ways one can 
generate four distinct constant 
coupling backgrounds in type IIB theory (see \cite{senF, DMF} for details).
Out of the four allowed backgrounds associated with the global symmetries ($D_4$, $E_6$, $E_7$ and  $E_8$), 
two -- $D_4$ and $E_7$ -- can be pulled to the heterotic side using our technique without generating extra fluxes.
The final metrics for the two cases will look exactly the same as in \eqref{hetbag} except that 
$\varphi_3$ will be periodic with period
\begin{equation}\label{period}
 0 ~\le ~\varphi_3 ~\le ~\pi ~~{\rm for}~~D_4 ~~~~{\rm and} ~~~~  0~ \le ~ \varphi_3 ~ \le~ \frac{\pi}{2} 
~~ {\rm for}~~ E_7,
\end{equation}
and the resulting picture will be a configuration of heterotic five-branes associated with these global symmetries.
For the other two cases, namely global symmetries $E_6$ and $E_8$, the heterotic backgrounds are defined 
with
\begin{equation}\label{e6e8}
 0 ~ \le ~ \varphi_3~ \le ~\frac{2\pi}{3} 
~~{\rm for}~~E_6 ~~~~{\rm and} ~~~~  0~ \le ~\varphi_3 ~\le~ \frac{\pi}{3} ~~ {\rm for}~~ E_8,
\end{equation}
and with an additional two-form NS flux $B_{\rm NS} = {1\over 2} d\varphi_1 \wedge d\varphi_2$ with zero field strength.
This flux is not globally defined because the $\varphi_i$ are not globally defined coordinates. Furthermore the dilaton for
these two cases differs slightly from \eqref{hetbag} by an overall constant: 
\ba
\label{dile8}
e^\phi = \frac{\sqrt{3}}{4}\left(R^2 ~{\rm sin}^2 \gamma ~{\rm sin}~2\psi\right) \, ,
\ea
leading to another configuration of heterotic five
branes.\footnote{It is interesting to note that the Argyres-Douglas points 
\cite{argdoug} would correspond to the following periodicities for $\varphi_3$: $\left[0, {5\pi \over 3}\right]$, 
$\left[0, {3\pi \over 2}\right]$, and $\left[0, {4\pi \over 3}\right]$, associated respectively with 
$A_0, A_1$ and $A_2$ global symmetries in type IIB theory. 
An analysis of the metric, dilaton and torsion can easily be done for these
cases too but we will not dwell on them here as our emphasis is more on studying axions than new non-K\"ahler 
manifolds.} Vector bundles in this set-up correspond to an intersecting configuration of seven branes, much like 
the one studied by Gimon and Polchinski \cite{gimon}. A more detailed discussion of these issues will be addressed elsewhere.

\subsubsection{Related AdS-type backgrounds}
Before we end this section, 
we note that if we change the orientation of the three-form flux $H$ from $\varphi_1, \varphi_2$
to $\varphi_1, \varphi_3$, keeping other factors unchanged, we can generate a slightly different background that falls
in the same class as \eqref{hetbag}:
\begin{eqnarray}\label{hetbagtwo}
\nonumber  ds^2 & =&  {1\over 2} R^2 ~ {\rm sin}~2\gamma ~{\rm cos}~\psi \Big[{r^2 \over R^2} 
 dx_{\mu} dx^{\mu}+ {R^2 \over r^2} ~dr^2\Big]  \\
 \nonumber &&+ {1\over 2} R^4~ {\rm sin}~2\gamma ~{\rm cos}~\psi \left[d\gamma^2 + {\rm sin}^2 \gamma~d\psi^2 
+ {\rm sin}^2 \gamma~ {\rm sin}^2 \psi~
d\varphi_2^2\right]~  \\ \nonumber 
&&  +  {\rm tan}~\gamma ~ {\rm cos}~\psi~d\varphi_3^2 + 
{\rm cot}~\gamma ~ {\rm sec}~\psi ~d\varphi_1^2; \\
e^\phi &= & {1\over 2g_s}\left(R^2 ~{\rm sin}~2\gamma ~{\rm cos}~\psi\right);\\  \nonumber
H&=& -\frac{4 R^{4}}{g_s^2} \sin^{3} \gamma \cos \gamma \sin \psi \cos \psi \,    d \gamma \wedge  d\psi 
\wedge  d \varphi_{2}      + {\cal O}(\alpha')        \,  .
\end{eqnarray}
This metric also has singularities. The Ricci scalar for this background in the Einstein frame scales like
${\cal R}  \sim  \sin^{-5/2} \gamma \,   \cos^{-5/2} \gamma \,   \sin^{-5/2} \psi  $, which diverges at
$\gamma=0$, $\gamma=\pi/2$ and $\psi=\pi/2$. Excising these points, and cutting off the radial direction to replace it with a finite cap we can have a smooth non-K\"ahler manifold that has well-defined curvature forms. The full torsion for
the manifold to higher orders in $\alpha'$
can now be easily computed following our earlier analysis \eqref{tors}. 

Yet another background, also falling in the same class, can be derived by changing the orientation of the three-form
from $\varphi_1, \varphi_2$ to $\varphi_2, \varphi_3$. This corresponds to taking
$\sin \psi \rightarrow \cos \psi$ in \eqref{hetbagtwo}:
\begin{eqnarray}
\label{hetbagthree}
 \nonumber ds^2&  =&  {1\over 2} R^2 ~ {\rm sin}~2\gamma ~{\rm sin}~\psi \Big[{r^2 \over R^2} 
dx_{\mu} dx^{\mu}+ {R^2 \over r^2} ~dr^2\Big]  \\ \nonumber
&&+ {1\over 2} R^4~ {\rm sin}~2\gamma ~{\rm sin}~\psi \left[d\gamma^2 + {\rm sin}^2 \gamma~d\psi^2 
+ {\rm sin}^2 \gamma~ {\rm cos}^2 \psi~
d\varphi_1^2\right]~  \\ \nonumber 
&&+  {\rm tan}~\gamma ~ {\rm sin}~\psi~d\varphi_3^2 + 
{\rm cot}~\gamma ~ {\rm cosec}~\psi ~d\varphi_2^2; \\  
e^\phi & =& {1\over 2g_s}\left(R^2 ~{\rm sin}~2\gamma ~{\rm sin}~\psi\right);\\  \nonumber
H&=& -\frac{4 R^{4}}{g_s^2} \sin^{3} \gamma \cos \gamma \sin \psi \cos \psi \,    d \gamma \wedge  d\psi 
\wedge  d \varphi_{1}      + {\cal O}(\alpha')        \,  .
\end{eqnarray}
This geometry also has singularities at $\gamma=0$, $\gamma=\pi/2$ and $\psi=0$, and following the same procedure 
as before we can have a smooth non-K\"ahler geometry with torsion and non-trivial vector bundles.\footnote{Note however 
that there is an apparent
obstruction to pulling the constant coupling backgrounds of \cite{senF, DMF} to the heterotic side using 
our sigma-model identifications.}


\subsection{The Axion Decay Constant}

Having constructed specific warped AdS heterotic backgrounds we can calculate the normalization constant
$\beta$ from (\ref{beta}) and find the axion decay constant $f_{a}$ from (\ref{ourf}).
The AdS geometries presented in the previous section
should be considered as local warped regions
or throats, which are glued in the UV to the compactification bulk.
The warped throat is glued to the bulk at $r=L$, where for consistency we impose $R   \, \simeq L$  , where $ R = (4\pi g_{s} N)^{1/4}  $ is
the AdS curvature radius  of the AdS geometry.
The overall size of the bulk of the compactification, $R_{6}$, is assumed to be much bigger
than the size of the throat, $R_{6}\gg L$, such that the bulk contains most of the volume of the compactification. Furthermore, it is assumed that the bulk is not warped.

The AdS geometry is also subject to an IR cut off, when $r \rightarrow 0$.
There is a conical singularity at $r=0$ and we assume that the geometry near the tip
of the cone or throat is modified such that this singularity is smoothed out as in the 
Klebanov-Strassler (KS) background \cite{Klebanov:2000hb}. The IR geometry is cut off 
at $r=r_{0}$ and the value of the warp factor $h_0$ at $r_0$ (after integrating the angular directions)
 is given by
\ba
\label{h0}
h_{0} =\frac{r_{0}}{\sqrt R} \, .
\ea
As in the KS solution, there are corrections to $h_{0}$ due to IR modification of the throat.
We expect them to be sub-leading and that they will not play a significant role in our discussion.

Noting that $4 \pi \kappa_{10}^{2} = m_{s}^{-8}$ and defining the volume of the bulk to be 
$v_{6} = R_{6}^{6}/4\pi$, 
the four-dimensional gravitational coupling from (\ref{MP}) is
\ba
\label{MP2}
M_{P}^{2}&= &4\pi m_{s}^{8}  \left(
 \frac{R_{6}^{6}}{4\pi} + \frac{1}{2}\pi^{3} R^{4} \alpha'^{3} g_{s}^{\frac{3}{2}} (1-\sin^{4} \gamma_{1}) 
(\sin^{2}\psi_{2} -  \sin^{2}\psi_{1} )(L^{2} -r_{0}^{2})
 \right) \nonumber\\
&\simeq& m_{s}^{8} R_{6}^{6} \left(  1+ 2 \pi^{4}  g_{s}^{\frac{3}{2}}  \frac{R^{4}   L^{2} \alpha'^{3}  }{  R_{6}^{6}  }
\right) \nonumber\\
&\simeq&   m_{s}^{8} R_{6}^{6} \, .
\ea
Here $\gamma_{1}$ and $\psi_{i}$ are the cut-off values for the 
angular variables $\gamma$ and $\psi$ at the singular points (\ref{ncpoint}).
In going from the first line to the second line above, it is assumed that $r_{0}\ll L$, 
$\gamma_{1}\rightarrow 0, \, \psi_{1}\rightarrow 0$ and $\psi_{2}\rightarrow \pi/2$.
To obtain the final answer, 
as mentioned before, it is assumed that the bulk contains most of the volume of the compactification, corresponding
to $R^{4} L^{2} \alpha'^{3} \sim L^{6} \alpha'^{3} \ll R_{6}^{2}$.

Similarly, for $\beta$ we obtain
\ba
\beta &=& \left(\frac{R_{6}^{6}}{4\pi} \right)^{-1}
\left[  e^{-\phi_{B}}  \frac{R_{6}^{6}}{4\pi} +  4 \pi^{3} R^{4} \alpha'^{3}  g_{s}^{\frac{5}{2}}
\left| \ln\left(   \frac{ \tan \psi_{2}  }{\tan \psi_{1}   }  \right)  \ln (\sin \gamma_{1}  ) \right|
\left(\frac{1}{r_{0}^{2}} -   \frac{1}{L_{0}^{2}}\right)  \right] \nonumber\\
&\simeq&    \frac{g_{s}^{\frac{5}{2}} }{4\pi^{2}} \frac{m_{s}^{2}}{ M_P^{2} }
 \left| \ln\left(   \frac{ \tan \psi_{2}  }{\tan \psi_{1}   }  \right)  \ln (\sin \gamma_{1}  ) \right|
  \frac{R^{4}}{ r_{0}^{2}  }.
\ea
In the first line, it is assumed that in the bulk the dilaton field does not change significantly,
and we denote its value in the bulk by  $\phi_{B}$.
To go from the first line to the second line it is assumed that $r_{0}/R^{2}<<1$, so the second term
dominates over the first term. Physically, this means that the normalization of the zero mode of $B_{NS}$ gets its largest contribution from the highly warped throat \cite{Mukhopadhyaya:2002jn, Mukhopadhyaya:2007jn, Firouzjahi:2007dp}. This should be contrasted with the normalization of the graviton zero mode, which is insensitive to the warp factor \cite{Firouzjahi:2005qs}. The calculation of the four-dimensional gravitational coupling in (\ref{MP2}) reflects this.

Plugging this value of $\beta$ into (\ref{ourf}), we obtain the axion decay constant 
\ba
\label{ourf1}
f_{a} \simeq \frac{ {\sqrt {8}} \,  \pi}{g_{s}^{{5}/{4}} R^{3/2}}
 \left|     \ln\left(   \frac{ \tan \psi_{2}  }{\tan \psi_{1}   }  \right)  \ln (\sin \gamma_{1}  )  \right|^{-1/2}
 (h_{0} \,  m_{s} ) \, .
\ea
The dependence on the cut-off angles $\gamma_{1}$ and $\psi_{1,2}$ is expected on physical grounds: the cut-off represents the deformation of the geometry near the singularities. These deformations will eventually show up in $\beta$ and $f_{a}$, when integrals over the non-singular compactification are performed. However, the axion decay constant is very insensitive to the angular coordinate cut-off; it  depends only logarithmically on $\gamma_{1}$ and $\psi_{1,2}$. As long as one is not exponentially fine-tuning the cut-off parameters to their singular values, the logarithmic expressions in $f_{a}$ will be of ${\cal O}(1)$. On the other hand, $R= (4 \pi g_{s} N )^{1/4}$. For parameters of
physical interest, one can assume $g_{s} \sim 0.1$ and
$1\lesssim     g_{s} N \lesssim 100$ 
such that $R \gsim  1$. Combining all these in $f_{a}$, we obtain the following expression for the axion decay constant:
\ba
\label{ourf1b}
f_{a}  =   c\,   h_{0} m_{s} \, ,
\ea
where $c$ is a constant of order $1 - 10$, depending on the geometry of the throat. 
Recall that $h_{0} \, m_{s}$ is nothing but the physical mass scale at the bottom of the throat.
This indicates that the axion decay constant is controlled by the physical scale of the throat and is 
insensitive to the details of the bulk.

To obtain $f_{a}$ within the acceptable range, i.e. $10^{9}$ GeV $\lesssim f_{a}\lesssim 10^{12}$ GeV,
all one has to do is to construct a throat in the string theory compactification  with the physical
mass scale within this range. This is easily achieved in the light of recent progress in flux
compactifications \cite{sav, Giddings:2001yu}.

The situation becomes more interesting in a multi-throat compactification.
One immediate conclusion of the result above is that in the multi-throat compactification,
the normalization of the axion field (or $B_{\mu \nu}$ field) zero mode is controlled by the longest throat
in the compactification.
To obtain axion decay of the right scale, one has to make sure that the physical mass scale of the
longest throat is within the range $10^{9}- 10^{12} $ GeV. In the case where the physical scales
of the throats are comparable, our formulation for calculating  $\beta$ indicates that
\ba
\label{multi}
f_{a } =\left[ \sum_{i} c_{i}^{-2} h_{0 \, i}^{-2}  \right]^{-1/2}\,  m_{s}
\ea
where the sum is over all throats. Here
$h_{0\,  i}$ represents the warp factor at the bottom of the $i$-th throat and the
$c_{i}$ are constants of order unity or so depending on the construction of the corresponding throat. Thus, even in the situation where the physical mass scale for each throat is bigger than $10^{12}$ GeV, all throats contribute to the normalization of the axion
zero mode such that the sum in (\ref{multi}) can bring $f_a$ within the desired range.


\section{Constant Coupling Background}

The previous examples we have studied give rise to large $\beta$ provided we impose a reasonable cut-off when compactifying the geometry. Two important aspects of our previous analysis were firstly that the dilaton remained  independent of the radial coordinate $r$, even though the background had non-trivial torsion, and secondly, that the analysis of $f_a$ was insensitive to the cut-off. This was not surprising since local cut-offs in the geometry should not affect many of the global features of a system. However, it would be nice to construct a background with torsion in the 
heterotic theory that is compact from the beginning, and allows a dilaton that is (at least) independent of the radial coordinate $r$. 

In the following we will sketch a possible background, with torsion and a constant dilaton. We will argue that in this background $\beta$ as defined in (\ref{beta}) can be made significantly greater than one.


A brief discussion of this issue appeared in \cite{gttwo}, where new heterotic backgrounds found via M-theory were studied. Our analysis here will similarly use M-theory to arrive at the heterotic theory, and will therefore be slightly different from the analysis presented in Section \ref{AdS}. As discussed in \cite{gttwo}, we want to emphasise that the localized and non--localized G-fluxes in M-theory play an important role in determing the precise backreaction effects on the underlying geometry. The conclusion was that the heterotic manifold will always be non--K\"ahler when only non--localized M--theory fluxes are present. On the other hand, in the presence of only localized fluxes, 
the manifold may or may not be K\"ahler. For our case, we will take the following ansatz for the metric of
the heterotic theory in the Einstein frame:
\begin{equation}\label{hetM}
ds^2 = e^{\cal D} g_{\mu\nu} dx^\mu dx^\nu + g_{mn} dx^m dx^n,
\end{equation}
where ${\cal D}$ is the warp factor, 
$\mu, \nu = 0,..., 3$ and $m,n$ denote internal coordinates of the manifold that may or may not be K\"ahler to
begin with.
The
generic formula for the non--K\"ahlerity is given in terms of the torsion classes ${\cal W}_i$ in the following 
way:\footnote{See \cite{lust,louis} for details about torsion classes.}
\begin{equation}\label{torsionclass}
dJ ~ = ~ \frac{3i}{4}\left({\cal W}_1 \bar\Omega - \bar{\cal W}_1 \Omega\right) + {\cal W}_3 + J \wedge {\cal W}_4,
\end{equation}
where $\Omega$ is the (3,0)-form for the manifold. If we demand that the internal manifold be complex then 
${\cal W}_1  = {\cal W}_2 = 0$. For our case we will also demand that $dB = 0$ 
and vanishing cosmological constant. This means that 
the torsion will only come from 
the Chern-Simons term and we have a four-dimensional Minkowski space (see 
\cite{frey} for a discussion of cosmological constant from the torsion classes). The general formula for non-K\"ahlerity can then be found from \eqref{torsionclass} to take the 
following form:
\begin{equation}\label{nonkah}
dJ ~ = ~ \alpha'\ast\left[\Omega_3(\omega_+)-\Omega_3(A)\right]
+ a_1~ e^{-2\phi} d\phi \wedge J + a_2~ d{\cal D} \wedge J,
\end{equation}
where $a_1$ and $a_2$ are constant coefficients whose values could be determined from \eqref{torsionclass},
$A$ is the one-form gauge bundle and $\phi$ is the heterotic dilaton. Observe that we have not identified the dilaton 
$\phi$ with the warp factor ${\cal D}$. This is therefore more generic than the corresponding formula presented in \cite{hull,rstrom,smit}.

Thus, even in the absence of a background three--form, the manifold can become non--K\"ahler due to the presence of vector bundles, a non-trivial
dilaton and the warp factor. K\"ahlerity is restored only when
\begin{equation}\label{kaha}
\omega_+ ~ = ~ A, ~~~~~~ \phi ~ = ~ {\rm constant}, ~~~~~~{\cal D} = 0,
\end{equation}
which are precisely the conditions studied in \cite{chsw}. Here we have derived the
conditions by demanding a K\"ahler compactification from the generic
equation for $dJ$.


Let us now imagine that we do not turn on any non--localized gauge
fluxes and at the same time do not allow the standard embedding. 
Then naively we would expect to get a
non--K\"ahler manifold with the non--K\"ahlerity coming precisely
from the difference $\Omega_3(\omega_+) - \Omega_3(A)$ (and the dilaton and warp factor) where
\begin{equation}\label{omega3}
\Omega_3(A) ~=~ \frac{1}{30}{\rm Tr}\left(A \wedge F - \frac{1}{3} A \wedge A \wedge A\right),
\end{equation}
and $\Omega_3(\omega_+)$  is defined equivalently but with a trace in the fundamental representation of the holonomy 
group. 
In fact, the three--form fluxes will typically look like \cite{bbdp, bbdgs}
\begin{equation}\label{thrform}
H_{ABC} ~ = ~ f_{ABC} + {\alpha'\over 2} ~{\rm Tr}~\Big( \omega  \wedge
\tilde f \wedge \tilde f + \tilde f \wedge {\cal R}_{\omega} +
{1\over 2} \tilde f \wedge d\tilde f  - {1\over 6} \tilde f \wedge \tilde f
\wedge \tilde f\Big)_{ABC} + {\cal O}(\alpha'^2),
\end{equation}
where $f_{ABC} = \alpha'[\Omega_3(\omega) - \Omega_3(A)]_{ABC}$ and 
$\omega$ is the gravitational
spin connection at zeroth order in $\alpha'$ (see \cite{bbdgs} for
more details). We have also defined $\tilde f$ as a one--form created from
$f$ using the vielbeins i.e ${\tilde f}_A^{~~ab} \equiv  f_{ABC} e^{Ba} e^{Cb}$ where $e^{Aa}$ is the vielbein, 
and ${\cal R}_\omega$ is defined in \eqref{romega}. In particular the traces are defined in the following way:
\begin{equation}\label{trace}
{\rm Tr} \left(\omega  \wedge \tilde f \wedge \tilde f\right)_{ABC} \equiv \omega_{[A}^{~~ab}~ {\tilde f}_{B}^{~~cd} 
~{\tilde f}_{C]}^{~~ef} ~{\rm Tr} \left(M_{ab} M_{cd} M_{ef}\right),
\end{equation}
where $M_{ab}$ is the matrix representation of the holonomy group. Similar definitions work for the other 
terms in $H_{ABC}$.
 
The above three--form backreacts on the geometry to make the space non--K\"ahler.
Thus, vector bundles without the standard embedding seem to be
allowed only on non--K\"ahler manifolds, unless there is an additional $\alpha'$ correction that could cancel the 
Chern-Simons term in \eqref{nonkah}.
At this point it would be interesting to compare the result with a similar construction 
discussed in \cite{Kgukov} where a
fractional gauge Chern--Simons term was switched on the internal manifold.
The key additional ingredient was 
a background {gaugino condensate}.\footnote{The gaugino condensate contributes to the (3,0) and the
(0,3) parts of the three--form, and can break susy, but we will consider a condensate that 
preserves susy.}

To conclude, the manifold can still become
non--K\"ahler in the absence of flux via the relation \eqref{nonkah}.
In \cite{Kgukov} the $\Omega_3(\omega)$ term was
cancelled by one of the Chern-Simons terms of the gauge fields.
Therefore the non--K\"ahlerity in this model arose from
vector bundles of one of the $E_8$ gauge groups. To obtain a Calabi--Yau
space we need $dJ = 0$. For this to be possible we would need to cancel the Chern-Simons term with the condensate. Fortunately the condensate is of the same order in $\alpha'$ as the Chern-Simons term. 
Furthermore, the presence of the condensate provides 
an additional contribution to the superpotential of \cite{bbdp, lustu}.
This additional contribution 
was worked out in \cite{lustd, bbdp} following the
work of \cite{gcon} (see also \cite{lilia} for some recent works). Additionally, the ten-dimensional Lagrangian for the heterotic theory will change from \eqref{10D} to the 
following:
\begin{eqnarray}\label{10Dnow}
S &= &  \frac{1}{ 2 \kappa^{2}} \int d^{10} x\,  \sqrt{-g}\Bigg[ R - \frac{1}{2} \partial_{M} \phi \partial^{M} \phi-
\frac{1}{12} e^{-\phi}  \left(H_{ABC} - \frac{\alpha'}{16}e^{\phi\over 2} 
\bar\chi \Gamma_{ABC} \chi\right)^2 \\ \nonumber
&& ~~~~~~~~ -\frac{\alpha'}{120} e^{-\frac{\phi}{2}}  \, {\rm Tr} \left(F_{AB} F^{AB}\right) - 
\frac{\alpha'}{30} {\rm Tr}\left(\bar\chi \Gamma^A D_A \chi\right)\Bigg],
\end{eqnarray}
where the $\chi^A$ are the gaugino fields of the heterotic theory. From here one can see that the 
equation for non--K\"ahlerity is now given by
\begin{equation}\label{nonnow}
dJ ~ =~ \alpha'\ast\left[\Omega_3(\omega_+)-\Omega_3(A)\right]
+ a_1~e^{-2\phi} d\phi \wedge J + a_2~d{\cal D} \wedge J + 
a_3~ \alpha'\ast\langle{\bar\chi}^A \Gamma \chi^A\rangle.
\end{equation}
The above equation allows for the following scenarios:

\noindent $\bullet$ Cancelling the Chern-Simons term with the gaugino condensate\footnote{Of course the Chern-Simons
term has to be of the form (3,0) $+$ (0,3) with no (2,1) $+$ (1,2) part. This is highly dependent on the 
complex structure, and one example of this is provided in \cite{Kgukov}.} 
and the dilaton term with the 
warp factor would lead to a K\"ahler manifold. With the underlying $SU(3)$ structure this would eventually 
become a Calabi-Yau manifold. 

\noindent $\bullet$ The cancellation of 
Chern-Simons term with the condensate is not always possible. However the dilaton 
and warp factor can be both made proportional to a constant. Then the manifold is K\"ahler to the zeroth order in 
$\alpha'$ and becomes non-K\"ahler to the first order in $\alpha'$ i.e. $dJ = {\cal O}(\alpha')$. 
It would be interesting to see whether the 
construction of \cite{Kgukov} falls in this category or the previous one. 
  
\noindent $\bullet$ The Chern-Simons term is cancelled by the condensate, but the dilaton is a constant. Then the 
manifold is non-K\"ahler due to the warp factor i.e. $dJ = {\cal O}(1)$. 
When the Chern-Simons term is not cancelled but the dilaton is
still a constant then the manifold remains non-K\"ahler i.e. $dJ = {\cal O}(1) + {\cal O}(\alpha')$.

\noindent $\bullet$ Taking the contribution of the background three-form \eqref{thrform}, one might be able
to retain K\"ahlerity to zeroth and first order in $\alpha'$. This would mean that the manifold could become 
non-K\"ahler at second order in $\alpha'$ i.e. $dJ = {\cal O}(\alpha^{'2})$. Of course non-K\"ahlerity to 
zeroth and first order in $\alpha'$ is still possible.

In this paper we will only discuss the implication of the third case, i.e. heterotic compactification with a constant 
dilaton, non-trivial warp factor and a torsion generated by the Chern-Simons term only (a gaugino condensate would 
make the system more involved). The background is therefore 
\begin{eqnarray}\label{torbag}
\nonumber  ds^2 ~&=&~ e^{\cal D} dx_{\mu} dx^\mu + g_{mn} dx^m dx^n; \\
 e^\phi ~& =& ~ {\rm constant};\\ 
 \nonumber   H ~ &=& ~ \frac{\alpha'}{120} ~{\rm Tr}~\left({1\over 3} A_1 \wedge A_1 \wedge A_1 - A_1 \wedge F_1\right),
\end{eqnarray}
where $A_1$ is the gauge field of one $E_8$ and the spin connection $\omega_+$ is embedded in the 
gauge connection $A_2$ of another $E_8$, i.e. $\omega_+ = A_2$. These manifolds are called {\it balanced} 
manifolds.\footnote{We thank Evgeny Buchbinder for pointing this out to us.}

The axion decay constant for this background is formally given by (\ref{ourf}) where $\beta$
is calculated from (\ref{beta}) with the replacement $h_{w}^{2}   \rightarrow   e^{\cal D}   $.
To find $f_{a}$ explicitly, we would need to know  $e^{\cal D} $ and $g_{mn}$ explicitly. 
However, motivated by our $AdS_{5}$ example, we believe that $f_{a}$ for this background can also
be lowered to within the narrow phenomenological bound. As explained before, the key property
here is that we are able to construct a background, at least in principle, where the dilaton
is independent of the warp factor. It would be interesting to work out the details of the supergravity
equations for this background and verify explicitly that $f_{a}$ can fall to within the desired window.

\section{Conclusion }
In the present work, the viability of axion constructions in warped heterotic backgrounds was studied. Manifolds considered in this context previously have in general been found to result in axion decay constants outside the allowed range, motivating the construction and investigation of new warped AdS-type heterotic backgrounds and their effect on the axion decay constant. As far as we know, these backgrounds, of the form (in string frame) 
\begin{equation}\label{lajbat}
ds^2 = e^\phi ds^2_{AdS_5} ~ + ~ ds^2_{X^5},
\end{equation}
with non-trivial dilaton and torsion, have not been studied before. From the four-dimensional point of view these backgrounds are warped Minkowski spacetimes with warp factor given by \eqref{wfac}. The compact six-dimensional manifolds are new non-K\"ahler manifolds, different from those studied in \cite{sav, beckerD, bbdg, GP, bbdgs}. This work can therefore be viewed in two ways: as a study of the effect of warping in string-theory backgrounds meeting certain criteria on the axion decay constant or as the phenomenologically motivated construction of new compact non-K\"ahler backgrounds in heterotic string theory. 

We gave one possible class of constructions of these backgrounds, which we believe to have smooth global metrics, using sigma-model identifications. It is interesting to note that the corresponding torsions can be expressed in powers of $\alpha'$ and satisfy the Bianchi identity in terms of the vector bundles and the torsional spin connections. We gave a simple way to construct vector bundles on these manifolds. A crucial feature of these backgrounds is that the dilaton is independent of the radial direction and therefore not quite proportional to the warp factor as in 
\cite{rstrom, sav, beckerD}.  A second class of non-K\"ahler manifolds with similar properties was also presented, with the important difference that the torsion comes only from the Chern classes of the manifolds. These manifolds have been classified in the literature as 
either  ``conformally balanced''  or ``balanced'' manifolds. Our case was an example of a balanced 
manifold with a constant dilaton but non-trivial torsion. 

We argued that  the zero modes of the 
 ``model-dependent'' axions may also give values of $f_a$ in the desired range, provided we have the precise knowledge 
of the harmonic two-forms i.e. $h_2$ in the internal space. These harmonic forms should be peaked near the throat of 
our internal space. Unfortunately, so far we have no precise knowledge of these forms 
and therefore we cannot make any definative
statement here. We believe that one may be able to find the required internal manifolds with harmonic 
forms satisfying the required properties. 

The axion in our construction is instead a ``model-independent'' axion, constructed from
the zero mode of the NS-NS two-form potential $B_{\mu \nu}$.  We have shown that the normalization of the 
zero mode of the $B_{\mu \nu}$ field gets most of its contribution from the highly warped regions of the 
compactification. It is on this fact that our mechanism for lowering the axion decay constant $f_a$ to the desired range ($10^{9}- 10^{12} $ GeV) hangs. More specifically, we have shown that $f_{a}$ is given by the warped mass scale of the longest throat
in the compactification. Hence, the question of achieving an axion construction with a viable value of $f_a$ in this set-up is reduced to the question of constructing an AdS-type throat with mass scale in the desired $10^{9}-10^{12}$ GeV range. 
As we explicitly demonstrated, 
heterotic compactification on any of the presented manifolds with appropriate torsion, dilaton and vector bundles should thus allow axions with permissible decay constants in a natural way.

Finally, one subtle issue that we have not addressed is the location of the standard model in our scenario. As is well known, there are two distinct ways we can argue for
the generation of the standard model spectrum. The first is the well known dimensional reduction of the $SO(32)$ or 
$E_8 \times E_8$ gauge fields on the non-K\"ahler manifold. Using the metric ansatze \eqref{metric} we see that the 
kinetic term of the gauge fields is {\it not} supressed by the warp factor $h_w$. However the interaction of the model-independent axion with the gauge fields is given by
\begin{equation}\label{csterm}
\frac{\alpha'}{4\kappa^2} \int e^{-\phi} ~dB \wedge \ast \Omega_3(A),
\end{equation}
where $\Omega_3(A)$ is the Chern-Simons term for the gauge field, and this is suppressed exactly as in \eqref{SH0}. Thus the
coupling of the axion (from the $B$ field) and the standard model gauge fields (from $\Omega_3(A)$) is suppressed by the warp factor, and therefore the coupling scale is set by the scale of the throat. 

The second way of generating the standard model, which is rather unconventional, is to use heterotic NS5 branes
wrapped on  two-cycles of the internal space. In the $SO(32)$ heterotic theory the NS5 brane is the small instanton
configuration and therefore has a direct gauge-theory spectrum on its world volume \cite{wittensmall}. 
On the other hand for the 
$E_8 \times E_8$ heterotic theory the NS5 brane spectrum is a tensor multiplet \cite{ganorhanany}. 
Multiple NS5 branes could 
give us a non-abelian four dimensional gauge theory. Clearly such a theory could be localised at the throat of
the manifold by allowing only suitably localised harmonic forms in the internal manifold. 
The axion coupling would then also be given by the 
scale of the throat instead of the bulk. Details on this will be presented elsewhere.  

\vskip.7cm

\centerline{\bf Acknowledgements}

\noindent We would like to thank E. Buchbinder, J. Conlon,
A. Frey, S. Kachru, J. Polchinski, H. Tye and A. Vilenkin  for helpful comments and discussions. This work is supported in part by NSERC. The work of RG is supported in part by a Chalk-Rowles fellowship. 

\vskip.2in

\appendix

\section{The Einstein Equation}
Here we give Einstein's equations for our solution in more detail.
The Einstein equation is  $G_{MN}=  \frac{1}{2}T_{MN}$ where the stress energy tensor is given by (\ref{Tmn}):
$$
T_{MN}=  \partial_{M} \phi \,   \partial_{N} \phi  
- \frac{1}{2} g_{MN} \partial_{P} \phi  \,  \partial^{P} \phi 
+ \frac{e^{-\phi} }{2 }  H_{MPQ} H_{N}^{\, \, PQ }
-\frac{e^{-\phi}}{ 12}   g_{MN}   H^2.
$$
To be specific, we calculate the components of Einstein tensor for the background (\ref{Ein1}):
\ba 
\nonumber && G_{00} ~ = ~ - G_{ii} =   
 \frac{ r^{2} \, (  1+1 2 \sin^{2} \psi \cos^{2} \psi \sin^{2} \gamma  )  }{ 4 \sqrt{g_{s}}  \alpha'   R^{2}\, 
 (\sin^{2} \gamma \cos^{2} \psi \sin^{2} \psi )    }; \\
\nonumber && G_{rr} ~ = ~ - \frac{R^{4}}{ r^{4}  }  \sqrt{g_{s}}  \alpha' \,      G_{00}; \\
\nonumber && G_{\gamma \gamma} ~ = ~
  \frac{  -1- 1 2 \sin^{2} \psi \cos^{2} \psi \cos^{2} \gamma   + 20  \sin^{2} \psi \cos^{2} \psi    }{4  \sin^{2} \gamma \cos^{2} \psi \sin^{2} \psi     }; \\ 
 &&  G_{\psi \psi} ~ = ~ (8- G_{\gamma \gamma} ) \sin^2 \gamma;\\
 \nonumber && G_{\phi1 \phi1}~ = ~ \frac{ \sqrt{g_{s}}  \alpha'   G_{ 00  } }{  r^{2} \sin^{2} \gamma \cos^{2}  \psi   }; \\
  \nonumber && G_{\phi2 \phi2}~ = ~ \frac{ \sqrt{g_{s}}  \alpha'  G_{ 00  } }{  r^{2} \sin^{2} \gamma \sin^{2}  \psi   }; \\
  \nonumber && G_{\phi3 \phi3}~ = ~ \frac{ \cos^{2} \gamma \,  ( -1+20 \sin^{2} \psi \cos^{2} \psi \sin^{2} \gamma )   }{4  \sin^{2} \gamma \cos^{2} \psi \sin^{2} \psi     };\\
   \nonumber && G_{\gamma \psi}~ = ~ \frac{ \cos \gamma(   \cos^{2} \psi -\sin^{2} \psi )  }{ \sin \gamma \sin \psi \cos \psi   }.
\ea
One can check that the off-diagonal component of the Einstein tensor is sourced by
$\partial_{\psi} \phi \partial_{\gamma} \phi$.
With these values for $G_{MN}$ along with $H$ and $\phi$ given as in (\ref{hetbag}), one can explicitly check that the Einstein equations are all satisfied. This demonstrates that the background  (\ref{Ein1})  is a genuine solution.
Similarly, one can also check that backgrounds (\ref{hetbagtwo}) and (\ref{hetbagthree}), when written in
the Einstein frame, are also consistent solutions with the given forms of $\phi$ and $H$.




\end{document}